\documentclass[journal]{IEEEtran}

\usepackage{color}
\usepackage{threeparttable}
\usepackage{srcltx}

\usepackage{graphicx,amsmath,psfrag,bm}
\usepackage[usenames,dvipsnames]{xcolor}
\usepackage{subfig}
\usepackage{cancel}
\usepackage{epstopdf}
\usepackage{pstool,cite}

\newcommand{\ve}[1]{\boldsymbol{#1}}
\newcommand{\te}[1]{\overline{\overline{#1}}}

\begin{document}

\title{Metasurface ``Solar Sail'' for flexible\\ Radiation Pressure Control}

\author{Karim~Achouri
        and~Christophe~Caloz
        }

\maketitle

\begin{abstract}
We propose to use metasurfaces as a mean of controlling radiation pressure for increasing the range of motions of spacecraft solar sails. Specifically, we present a theoretical study of different electromagnetic field configurations, and corresponding metasurface structures, that allow one to achieve repulsive, attractive, lateral and rotational forces.
\end{abstract}


\IEEEpeerreviewmaketitle

\section{Introduction}

The solar sail is a spacecraft propulsion method based on radiation pressure. Although the force density exerted by light upon scattering on an object is very small, the resulting force may be sufficient for propulsion if the scattering area is sufficiently large. This technology may, one day, allow humanity to travel among the stars~\cite{Johnson2011}. However, solar sails are, as of now, restricted to \emph{repulsive} forces, which severely limits the spacecraft motion capability.

In this work, we extend the range of operation of conventional solar sails by introducing metasurface solar sails. We propose to leverage the electromagnetic transformation capabilities of metasurfaces to control radiation pressure. While most studies on optical forces have been so far restricted to the manipulation of forces acting on small particles~\cite{Shvedov2014,Chen2011,Wang2014,PhysRevB.91.115408,Salary2016}, our goal here is to design a metasurface system, which consists of a metasurface attached to an object to be moved (e.g. a satellite), and whose motion can be controlled by the illumination emerging either from stars or from high-power earth-based or satellite-born lasers. Different forces may then be obtained by varying the polarization and/or wavelength of the illumination. In what follows, we propose a prospective study on the capabilities of metasurfaces to control radiation pressure.

\section{Electromagnetic Force on a Stationary Object}

\begin{figure}[htbp]
\begin{center}
\includegraphics[width=1\columnwidth]{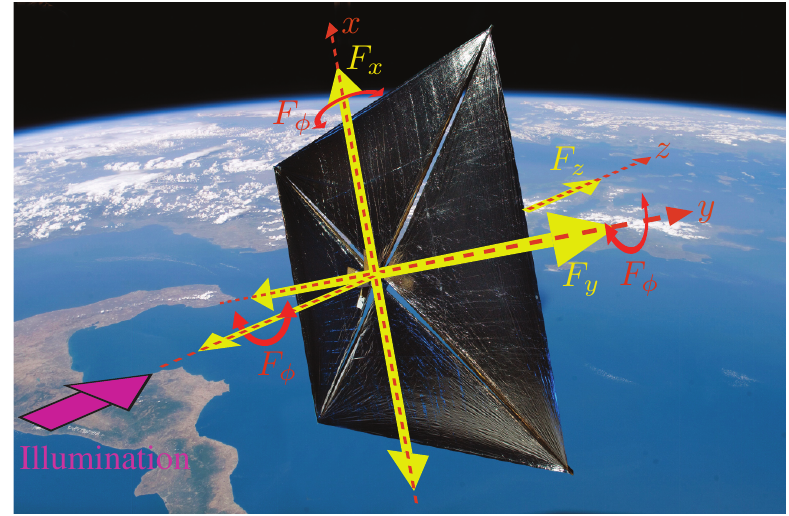}
\caption{Proposed metasurface solar sail with two lateral forces ($\pm F_x$ and $\pm F_y$), a repulsive/attractive force ($\pm F_z$) and three rotational forces ($\pm F_\phi$). Picture credit: NASA.}
\label{Fig:SS}
\end{center}
\end{figure}

An electromagnetic wave carries both energy and momentum. When it is scattered or absorbed by an object, the latter is subjected to a force as a consequence of the conservation of momentum law, which reads~\cite{rothwell2008electromagnetics}
\begin{equation}
\label{eq:momentumconservation}
\ve{f} + \epsilon\mu\frac{\partial \ve{S}}{\partial t} = \nabla\cdot\te{T}_\text{em},
\end{equation}
where $\ve{f}$ is the volume force density, $\ve{S}$ is the Poynting vector and $\te{T}_\text{em}$ is the Maxwell stress tensor, which is itself given by
\begin{equation}
\label{eq:MST}
\te{T}_\text{em}=\ve{D}\ve{E} + \ve{B}\ve{H}  - \frac{1}{2} \te{I}(\ve{D}\cdot\ve{E}+\ve{B}\cdot\ve{H}),
\end{equation}
where $\te{I}$ is the identity tensor and $\ve{E}, \ve{D}, \ve{B}$ and $\ve{H}$ correspond to the total electromagnetic fields.
Let us assume, for simplicity, that the object is not moving and hence that the total (incident and scattered) field around the object is not changing with time. In that case, the time derivative of the Poynting vector in~\eqref{eq:momentumconservation} vanishes. Using Gauss integration law, the time-averaged force acting on the object is then given by
\begin{equation}
\label{eq:TAF}
\langle \ve{F} \rangle =\int_V \nabla\cdot\langle\te{T}_\text{em}\rangle~dV= \oint_S \langle\te{T}_\text{em}\rangle\cdot\hat{n}~dS,
\end{equation}
where $\ve{\hat{n}}$ is the unit vector normal to the surface surrounding the object and $\langle \cdot \rangle$ denotes the time-average operation. Assume now that the object to be moved is the metasurface system surrounded by vacuum, as depicted in Fig.~\ref{Fig:SS}. The forces acting on this metasurface, which is located at $z=0$ in the $xy$-plane, are calculated using~\eqref{eq:TAF}. The surface integration in~\eqref{eq:TAF} is performed on two planar surfaces, which are located at $z=0^+$ and $z=0^-$ and for which ${\hat{n}}=+{\hat{z}}$ and ${\hat{n}}=-{\hat{z}}$, respectively.

Let us assume, for simplicity, that the interactions between the incident, reflected and transmitted waves and the metasurface occur only in the $xz$-plane. The metasurface is isotropic, has a finite lateral size of dimensions $L_x\times L_y$ and is assumed to be of zero thickness. We now calculate the force that a $p$-polarized plane wave, impinging at an angle $\theta_\text{i}$ from broadside, exerts on the metasurface. The corresponding electromagnetic fields of the incident wave are given by 
\begin{subequations}
\label{eq:fields}
\begin{equation}
\ve{H}_\text{i} = \hat{y}\frac{E_0}{\eta_0}e^{-jk_0(\sin{(\theta_\text{i})}x + \cos{(\theta_\text{i})}z)},
\end{equation}
\begin{equation}
\ve{E}_\text{i} = [\hat{x}\cos{(\theta_\text{i})}-\hat{z}\sin{(\theta_\text{i})}]E_0e^{-jk_0(\sin{(\theta_\text{i})}x + \cos{(\theta_\text{i})}z)},
\end{equation}
\end{subequations}
where $E_0$ is the amplitude of the wave. Inserting these fields in the time-averaged version of the Maxwell stress tensor in~\eqref{eq:MST} and post-multiplying by $\hat{n} = -\hat{z}$ to extract the components on a plane parallel to that of the metasurface at $z=0^-$ leads to
\begin{equation}
\label{eq:MST2}
\langle\te{T}_\text{em}\rangle\cdot\hat{n}= \hat{x}\frac{1}{2}E_0^2\epsilon_0\cos{(\theta_\text{i})}\sin{(\theta_\text{i})} + \hat{z}\frac{1}{2}E_0^2\epsilon_0\cos^2{(\theta_\text{i})}.
\end{equation}
Finally, the time-averaged forces are found by performing the integral in~\eqref{eq:TAF} over the area $L_x\times L_y$ of the metasurface, which yields
\begin{subequations}
\label{eq:FPW}
\begin{align}
\langle F_x \rangle^\text{i}&= \frac{1}{2}\epsilon_0E_0^2L_xL_y\cos{(\theta_\text{i})}\sin{(\theta_\text{i})},\label{eq:FPWx}\\
\langle F_z \rangle^\text{i}&= \frac{1}{2}\epsilon_0E_0^2L_xL_y\cos^2{(\theta_\text{i})},\label{eq:FPWz}
\end{align}
\end{subequations}
Similarly, the forces due to the reflected and transmitted waves may be straightforwardly deduced from~\eqref{eq:FPW} to be $\langle F_x \rangle^\text{r} = - \langle F_x \rangle^\text{i}$ and $\langle F_z \rangle^\text{r} =  \langle F_z \rangle^\text{i}$, and $\langle F_x \rangle^\text{t} = - \langle F_x \rangle^\text{i}$ and $\langle F_z \rangle^\text{t} = - \langle F_z \rangle^\text{i}$.

In order to evaluate the forces in a more realistic scenario, we also consider the case of Gaussian illumination. Let us consider a 2D Gaussian beam with a Gaussian profile $E~\propto~e^{-\ve{r}^2/w}$, where $w$ is related to the half-power beamwidth, through $\text{HPBW} = 2\sqrt{w\ln(2)}$. The forces that are exerted on the metasurface are again found from~\eqref{eq:TAF} with~\eqref{eq:MST} and read
\begin{subequations}
\label{eq:2DGaussF}
\begin{align}
\label{eq:2DGaussFx}
&\langle F_x \rangle^\text{i}=\frac{E_0^2\epsilon_0L_y}{4k_0^2} \sqrt{w_\text{i}}\sin(\theta_\text{i}) \bigg[2L_x \sqrt{w_\text{i}}\cos(\theta_\text{i})e^{-L_x^2\cos(\theta_\text{i})^2/(2w_\text{i})}\\\nonumber
&\qquad\qquad+\sqrt{2\pi}(k_0^2-w_\text{i})\mbox{Erf}\Big(\frac{L_x\cos(\theta_\text{i})}{\sqrt{2 w_\text{i}}}\Big)\bigg],\\
\label{eq:2DGaussFz}&\langle F_z \rangle^\text{i}=\frac{E_0^2\epsilon_0L_y \sqrt{w_\text{i}}}{8k_0^2}\bigg[2L_x\sqrt{w_\text{i}}\cos(2\theta_\text{i})e^{-L_x^2\cos(\theta_\text{i})^2/(2w_\text{i})}\\\nonumber
&+\sqrt{2\pi}\big(k_0^2+(k_0^2-w_\text{i})\cos(2\theta_\text{i})\big)\sec(\theta_\text{i})\mbox{Erf}\Big(\frac{L_x\cos(\theta_\text{i})}{\sqrt{2 w_\text{i}}}\Big)\bigg],
\end{align}
\end{subequations}
where $\mbox{Erf}(x)$ is the error function. In the case of a 3D Gaussian illumination, the forces are directly found to be
\begin{subequations}
\label{eq:3DGaussF}
\begin{align}
\langle F_x \rangle_\text{3D}^\text{i}&=\sqrt{\frac{\pi}{2}}\text{Erf}\left(\frac{L_y}{\sqrt{2}}\right)\langle F_{x} \rangle_\text{2D}^\text{i},\label{eq:3DGaussFx}\\
\langle F_z \rangle_\text{3D}^\text{i}&=\sqrt{\frac{\pi}{2}}\text{Erf}\left(\frac{L_y}{\sqrt{2}}\right)\langle F_{z} \rangle_\text{2D}^\text{i},\label{eq:3DGaussFz}
\end{align}
\end{subequations}
where the terms with the subscripts ``2D'' refer to the forces in~\eqref{eq:2DGaussF}. These expressions provide the tools required to numerically investigate the different field configurations required to achieve the desired forces. From that point, we shall be able to synthesize the metasurfaces so as to realize these forces. This is the topic of the next section.

\section{Metasurface Mathematical Synthesis}

The mathematical synthesis of metasurfaces consists in obtaining the metasurfaces susceptibilities as functions of the specified electromagnetic transformations. The susceptibilities can be mathematically related to the specified incident, reflected and transmitted fields. Here, the metasurface synthesis is based on the technique previously developed by the authors in~\cite{achouri2014general,Achouri2015c}, which is itself based on the generalized sheet transition conditions (GSTCs)~\cite{Idemen1973}. In the case of a bianisotropic metasurface lying in the $xy$-plane at $z=0$,  the GSTCs read
\begin{subequations}
\label{eq:BC_plane}
\begin{align}
\hat{z}\times\Delta\ve{H}
&=j\omega\epsilon_0\te{\chi}_\text{ee}\cdot\ve{E}_\text{av}+jk_0\te{\chi}_\text{em}\cdot\ve{H}_\text{av},\label{eq:BC_plane_1}\\
\Delta\ve{E}\times\hat{z}
&=j\omega\mu_0 \te{\chi}_\text{mm}\cdot\ve{H}_\text{av}+jk_0\te{\chi}_\text{me}\cdot\ve{E}_\text{av},\label{eq:BC_plane_2}
\end{align}
\end{subequations}
where $\Delta$ indicates the difference of the fields between both sides of the metasuface, the subscripts ``av'' stand for the average of these fields and $\te{\chi}_\text{ee}, \te{\chi}_\text{mm}, \te{\chi}_\text{me}$ and $\te{\chi}_\text{em}$ are the electric, magnetic and electromagnetic susceptibility tensors, respectively.

The synthesis technique consists in specifying the desired electromagnetic transformation in terms of corresponding fields on both sides of the metasurface, and then solving~\eqref{eq:BC_plane} for the susceptibilities. Here, we will not further detail the synthesis procedure and refer the reader to~\cite{achouri2014general,Achouri2015c} for related in-depth discussions. Moreover, note that, in this theoretical work, we will only discuss the mathematical synthesis of some metasurfaces ``solar sails'' and leave the practical realization of these structures for a future work.

\section{Radiation Pressure Control with Metasurfaces}

\begin{figure}[ht!]
\begin{center}
\subfloat[]{\label{Fig:F1}
\includegraphics[width=0.45\columnwidth]{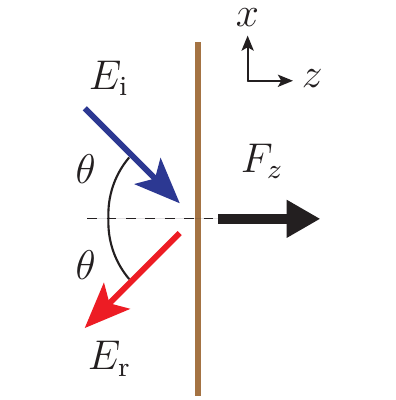}
}
\subfloat[]{\label{Fig:F2}
\includegraphics[width=0.45\columnwidth]{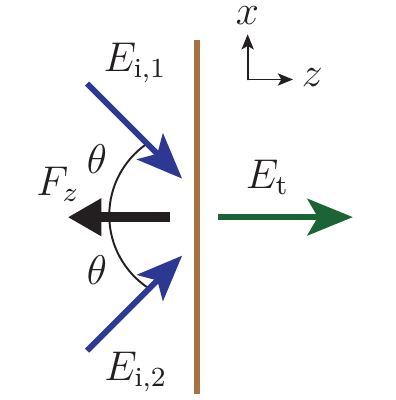}
}\\
\subfloat[]{\label{Fig:F3}
\includegraphics[width=0.45\columnwidth]{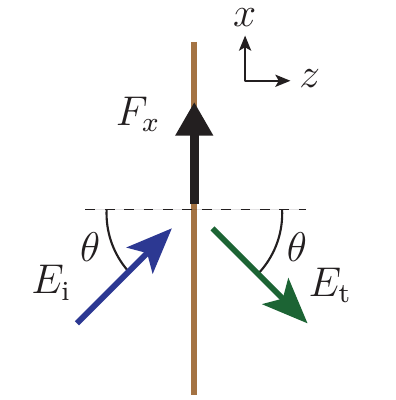}
}
\subfloat[]{\label{Fig:F4}
\includegraphics[width=0.45\columnwidth]{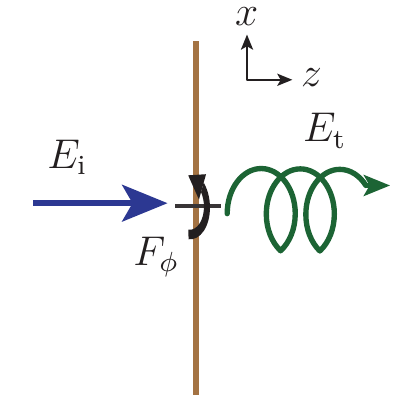}
}
\caption{Four field configurations corresponding to (a)~a repulsive force, (b)~an attractive force, (c)~a lateral force and (d)~an in-plane rotational force.}
\label{Fig:F}
\end{center}
\end{figure}

We are now interested in finding the incident, reflected and transmitted waves, acting on the metasurface system in Fig.~\ref{Fig:SS}, so as to generate repulsive, attractive, lateral and rotational forces. The four corresponding field configurations are represented in Figs.~\ref{Fig:F}, where they respectively correspond to the operations of specular reflection, wave combination, negative refraction and Bessel beam generation (for in-plane rotation). In what follows, we will investigate in more details the electromagnetic behavior of these different cases.

\subsection{Repulsive Force}

\begin{figure}[ht!]
\centering
\includegraphics[width=1\linewidth]{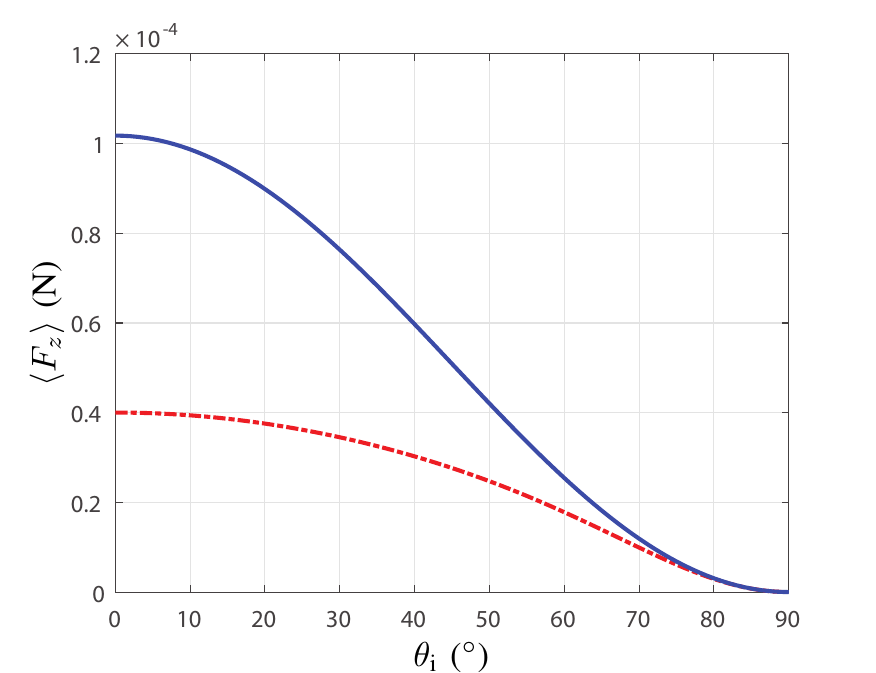}
\caption{Repulsive forces versus incidence angle exerted on a perfectly reflective surface assuming plane wave illumination (solid blue line) and Gaussian illumination (dashed red line) with $w_\text{i}=8$.}
\label{Fig:RepulsiveF}
\end{figure}

Achieving a repulsive force with a flat structure is rather easy. The maximal repulsive force is simply obtained with a perfectly reflective surface~\cite{Novotny2012}. Consider the illustration in Fig.~\ref{Fig:F1}, where an obliquely incident wave is specularly reflected with reflection coefficient $|R|=1$. In that case, the force acting on the object is only in the $z$-direction since the contributions of the incident and reflected waves along $x$ cancel each other. In the case of plane wave illumination, the longitudinal force is, from~\eqref{eq:FPWz}, given by
\begin{equation}
\langle F_z \rangle = \langle F_z \rangle^\text{i} + \langle F_z \rangle^\text{r} =  2\langle F_z \rangle^\text{i}= \epsilon_0E_0^2L_xL_y\cos{(\theta)}^2.
\end{equation}
The same procedure may be used to obtain the force due to a 2D Gaussian illumination, using~\eqref{eq:2DGaussFz}. Note that in that specific case of specular reflection, the beamwidth of the incident and reflected Gaussian beams are equal since $\theta_\text{i} = \theta_\text{r}$ and thus $w_\text{i} = w_\text{r}$. To illustrate the differences between the repulsive forces obtained with a plane wave illumination and with a Gaussian illumination, we plot these forces as functions of the incidence angle in Fig.~\ref{Fig:RepulsiveF} with the following parameters: $E_0=120\pi$ V/m, $L_x = L_y = 9$ m, $\lambda_0 = 500$ nm and $w_\text{i}=8$.
We see that both illuminations lead to the same force profile. As expected, the force due to the Gaussian illumination is smaller than that due to the plane wave illumination since less power is impinging on the metasurface in the former case. The maximum force is naturally obtained when the incident wave is normally impinging on the metasurface.

\subsection{Attractive Force}

A metasurface can be subjected to an attractive force if the incident waves are transformed into transmitted waves with momentum in the $z$-direction larger than that of the incident waves. This change of momentum results in a negative longitudinal force. One of the simplest situation is that of the reflectionless transformation depicted in Fig.~\ref{Fig:F2}. In this case, the metasurface combines two incident waves both impinging with opposite incidence angles along $x$. Here, we are considering two incident waves propagating with opposite $k_x$ wavenumber so that the lateral force, acting on the metasurface in the $x$-direction, is zero. Moreover, to maximize the attractive force, we specify that the metasurface is fully efficient, i.e. that all incident power is transmitted through the metasurface.

To achieve this specification, the two incident waves \emph{must} be orthogonally polarized and thus be treated independently from each other by the metasurface. Indeed, if the two incident waves had the same polarization, then the metasurface would act as a beam combiner or, if used in its reciprocal operation state, as a beam splitter. In that case, the efficiency would necessarily be limited to at best $50\%$. Therefore, the only way to efficiently realize the operation in Fig.~\ref{Fig:F2} is to consider two orthogonally polarized incident waves normally refracted by the metasurface. Finally, these two refraction operations can be realized with $100\%$ efficiency if the metasurface is bianisotropic, as extensively discussed in~\cite{Lavigne2017}. For conciseness, the metasurface susceptibilities are not provided here since they are exactly the same as those already provided in~\cite{Lavigne2017}.

In the case of a refracting metasruface, the beamwidth of the incident waves is smaller than that of the transmitted waves when $\theta_\text{i} > \theta_\text{t}$. Accordingly, the relation between these beamwidths is $w_\text{t}=w_\text{i}\cos^2(\theta_\text{t})/\cos^2(\theta_\text{i})$, while, to satisfy power conservation, the relation between the amplitude of the waves is $E_\text{t}=E_0\sqrt{\cos(\theta_\text{i})/\cos(\theta_\text{t})}$, where $E_0$ is the amplitude of the incident wave and $E_\text{t}$ that of the transmitted wave.

\begin{figure}[h!]
\centering
\includegraphics[width=1\linewidth]{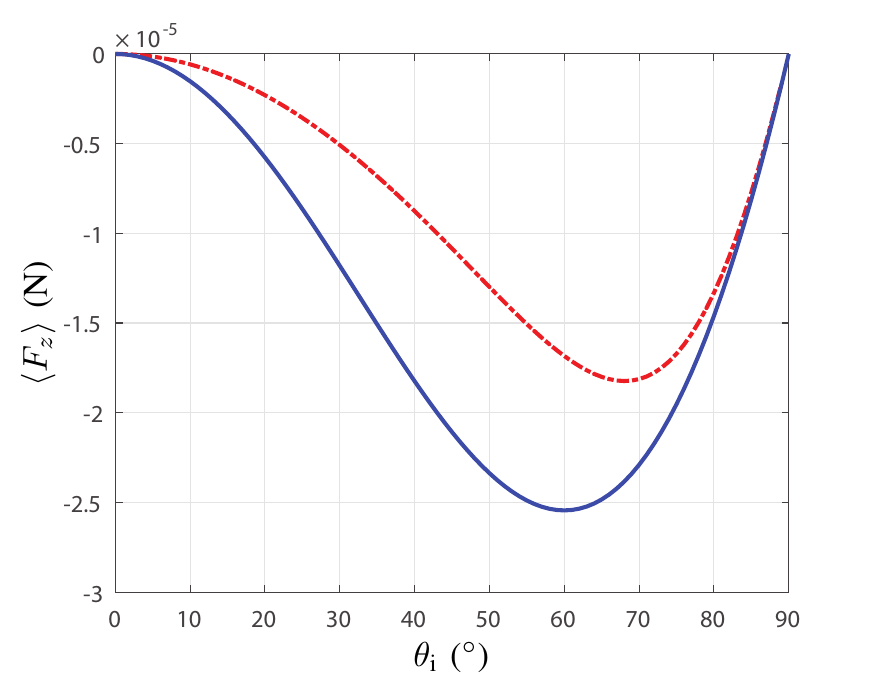}
\caption{Attractive forces versus incidence angle exerted on the metasurface assuming plane wave illumination (solid blue line) and Gaussian illumination (dashed red line) with $w_\text{i}=8$.}
\label{Fig:AttractiveF}
\end{figure}

Taking these considerations into account, the attractive force exerted on the bianisotropic metasurface by the two incident plane waves, as function of the incidence angle, with $\theta_\text{t} = 0^\circ$ for maximum force, is given by
\begin{equation}
\label{eq:Attrac}
\langle F_z \rangle = -2L_xL_y\epsilon_0E_0^2\cos{(\theta_\text{i})}\sin{\left(\frac{\theta_\text{i}}{2}\right)}^2.
\end{equation}
Interestingly, the maximum attractive force is achieved when $\theta_\text{t} = 0^\circ$ and $\theta_\text{i} = 60^\circ$. The corresponding attractive force for Gaussian illumination is found, using the same considerations as above, to be
\begin{equation}
\label{eq:AttracG}
\begin{split}
\langle F_z \rangle=&\frac{E_0^2\epsilon_0L_y\sqrt{w_\text{i}}\sec{(\theta_\text{i})}^2}{2k_0^2}\bigg[L_x\sqrt{w_\text{i}}\cos(\theta_{\text{i}})(\cos(\theta_{\text{i}})\cos(2\theta_{\text{i}})\\
&-1)e^{-L_x^2\cos(\theta_{\text{i}})^2/(2 w_\text{i})}+\sqrt{2\pi}\big(2w_\text{i}(1-\cos(\theta_{\text{i}}))-k_0^2\\
&+(k_0^2-w_\text{i})\cos(2\theta_{\text{i}})\big)\sin\left(\frac{\theta_{\text{i}}}{2}\right)^2\mbox{Erf}\Big(\frac{L_x\cos(\theta_{\text{i}})}{\sqrt{2 w_\text{i}}}\Big)\bigg].
\end{split}
\end{equation}
The relations~\eqref{eq:Attrac} and~\eqref{eq:AttracG} are compared in Fig.~\ref{Fig:AttractiveF} with the same parameters as before.
The maximum attractive force, in the case of Gaussian illumination, is shifted towards higher incidence angles compared to the case of plane wave illumination. This may be understood by first appreciating why the maximum force is at $\theta_\text{i} = 60^\circ$, in the plane wave illumination case. The force exerted on the metasurface is, by conservation of momentum, due to a change in the direction of wave propagation. Intuitively, the maximum attractive force should thus be obtained when $\theta_\text{i} = 90^\circ$ and $\theta_\text{t} = 0^\circ$. However, as $\theta_\text{i}$ increases, less and less energy is passing through the metasurface until, eventually, no power passes through when $\theta_\text{i} = 90^\circ$. It turns out that these two antagonist effects leads to a maximum attractive force at $\theta_\text{i} = 60^\circ$. Now let us consider Gaussian illumination. In that case, even for relatively large incidence angles, most of the incident power still remains within the limited surface area ($L_x\times L_y$) of the metasurface without spillover due to the Gaussian profile of this illumination. This effectively shifts the maximum force towards $\theta_\text{i} = 90^\circ$. Obviously, the smaller the beamwidth ($w_\text{i}$), the more confined is the incident power and thus the more important is the shift.

\begin{figure}[ht]
\centering
\includegraphics[width=1\linewidth]{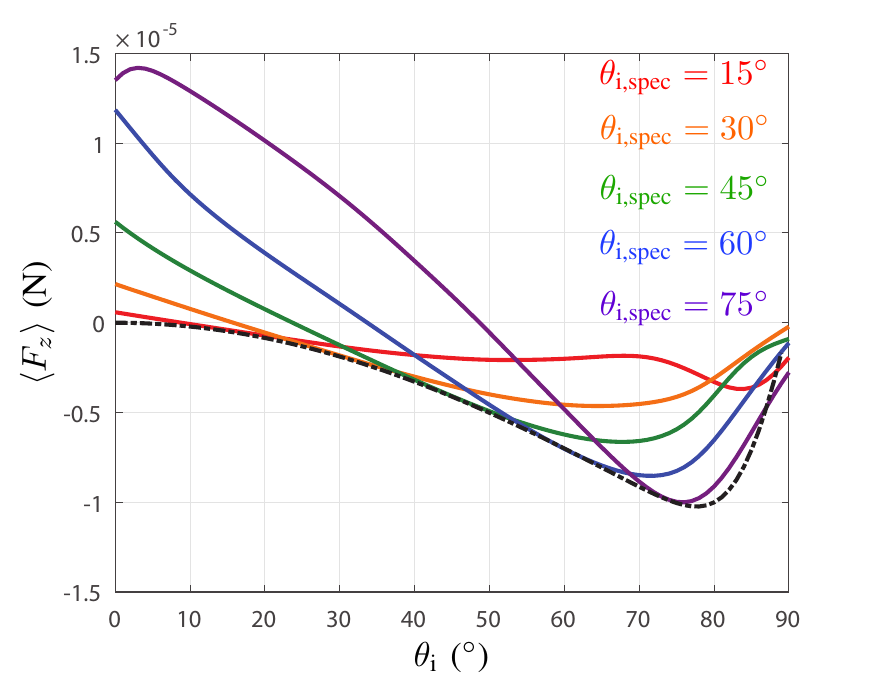}
\caption{Attractive forces versus incidence angle exerted on five different metasurfaces when the incidence angle deviates from the incidence angle specified in the synthesis. The dashed black line corresponds to the longitudinal force for the specified incidence angles; it is the same as the dashed red line plotted in Fig.~\ref{Fig:AttractiveF} but with $w_\text{i}=1$.}
\label{Fig:AttractiveFz}
\end{figure}

The plots in Fig.~\ref{Fig:AttractiveF} represent the forces exerted on the metasurface for \emph{specified} incidence angles. The question that arises now is: how would these forces be affected if the metasurface was illuminated by incident waves impinging at angles that are different from the specified incidence angle used to synthesize the metasurface? In order to evaluate this effect on the longitudinal force, we have performed 2D FDFD simulations, based on our previously developed technique~\cite{Vahabzadeh2016}, with five metasurfaces synthesized for the specified incidence angles $\theta_\text{i,spec} = \{15^\circ, 30^\circ, 45^\circ, 60^\circ, 75^\circ\}$. To maximize the refraction efficiency, the metasurfaces are bianisotropic, as mentioned above, and the corresponding susceptibilities are obtained from~\cite{Lavigne2017}. The simulation results are plotted in Fig.~\ref{Fig:AttractiveFz}. Note that the metasurfaces are illuminated with Gaussian illumination with a beamwidth that is smaller ($w_\text{i} = 1$) than that used in Fig.~\ref{Fig:AttractiveF} where $w_\text{i} = 8$.

As can be seen, the attractive force is more important when $\theta_\text{i,spec}$ is large, as expected. We can also see that the simulation results are in good agreement with the expected values at the points where $\theta_\text{i} = \theta_\text{i,spec}$ (corresponding to the dashed black line). Moreover, the acceptance angle, defined as the total angle variation from the specified incidence angle under which the metasurface is still subjected to an attractive force, is particularly important. This means that the generation of an attractive force, with such a metasurface, is robust to deviations from the specified illumination. Note that, to obtain the results in Fig.~\ref{Fig:AttractiveFz}, we have assumed that the incidence angles of the two incident waves are the same, meaning that the force along the $x$-direction is zero for all angles.

\subsection{Lateral Force}

\begin{figure}[ht!]
\centering
\includegraphics[width=1\linewidth]{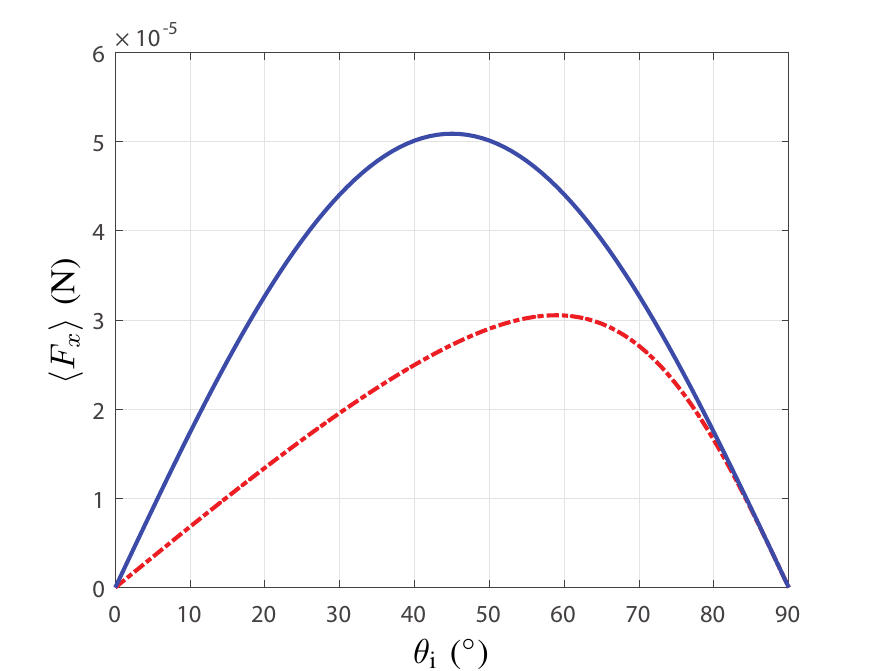}
\caption{Lateral forces versus incidence angle exerted on the metasurface assuming plane wave illumination (solid blue line) and Gaussian illumination (dashed red line) with $w_\text{i}=8$.}
\label{Fig:LateralF}
\end{figure}

The realization of a lateral force \emph{only} requires that the longitudinal momentum of the waves, on both sides of the metasurface, vanishes while the variation of momentum in the lateral direction (e.g. $x$-direction) is maximized. This may be achieved with the negative refractive transformation depicted in Fig.~\ref{Fig:F3}, where the incidence and transmission angles are equal to each other. Because these two angles are the same, the beamwidth as well as the amplitude of both incident and transmitted waves are the same. As before, the time-averaged forces acting on the metasurface are computed and the resulting longitudinal force is $\langle F_z \rangle = 0$ while the lateral force is, from~\eqref{eq:FPWx}, readily found to be
\begin{equation}
\begin{split}
\langle F_x \rangle &= \frac{1}{2}\epsilon_0E_0^2L_xL_y\cos{(\theta_\text{i})}\sin{(\theta_\text{i})} \\&
-\frac{1}{2}\epsilon_0E_0^2L_xL_y\cos{(\theta_\text{t})}\sin{(\theta_\text{t})} = \frac{1}{2}L_xL_y\epsilon_0 E_0^2\sin{(2\theta)},
\end{split}
\end{equation}
where $\theta_\text{t} = -\theta_\text{i}$. It is interesting to note that this force reaches a maximum for $\theta = 45^\circ$. The lateral force for Gaussian illumination can be obtained following the same procedure but with~\eqref{eq:2DGaussFx} and by setting $w_\text{t} = w_\text{i}$. The lateral forces acting on the metasurface for plane wave and Gaussian illuminations are plotted in Fig.~\ref{Fig:LateralF} versus specified incidence angles.
As was the case for the attractive force discussed above, we see that the maximum of the force, in the case of Gaussian illumination, is shifted towards larger incidence angles. The explanation for this effect is the same as the one given previously.

\begin{figure}[ht!]
\centering
\subfloat[]{\label{Fig:LateralFfx}
\includegraphics[width=1\linewidth]{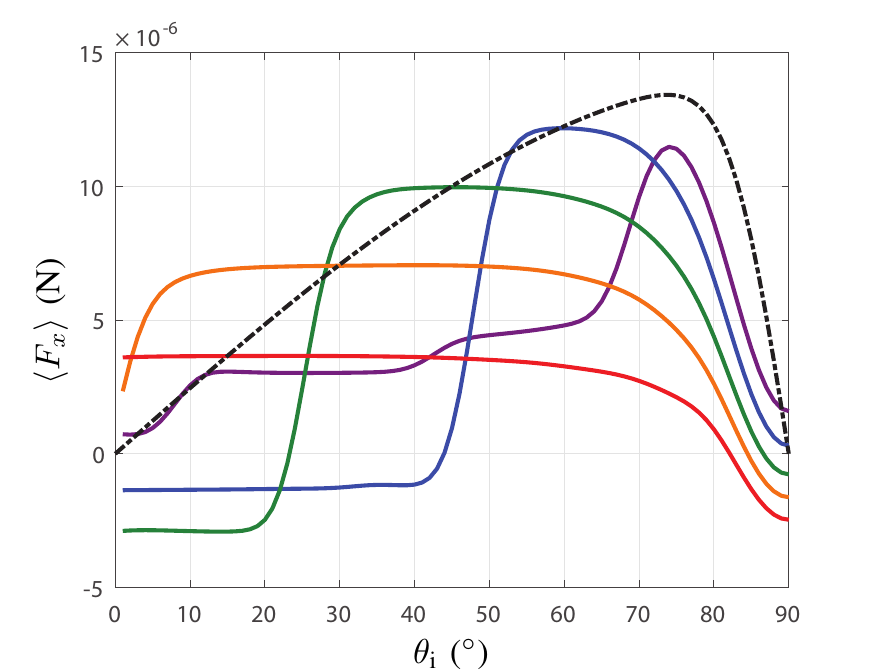}
}\\
\subfloat[]{\label{Fig:LateralFfz}
\includegraphics[width=1\linewidth]{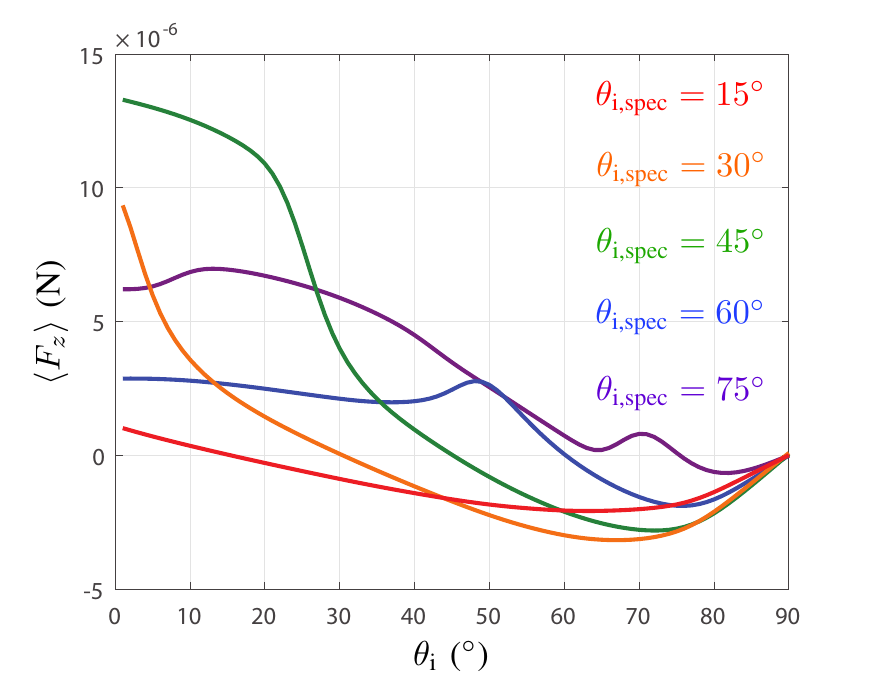}
}
\caption{Forces exerted on five different metasurfaces when the incidence angle deviates from the specified incidence angle used in the synthesis. (a) Lateral forces and (b) longitudinal forces. The dashed black line corresponds to the lateral force for specified incidence angles, it is the same as the one plotted in Fig.~\ref{Fig:LateralF} but with $w_\text{i}=1$.}
\label{Fig:LateralFf}
\end{figure}

Let us now evaluate the behavior of this metasurface when the incidence angle deviates from the specified one. We consider five different metasurfaces synthesized for the following specified incidence angles: : $\theta_\text{i,spec} = \{15^\circ, 30^\circ, 45^\circ, 60^\circ, 75^\circ\}$. The susceptibilities are obtained, assuming a monoisotropic metasurface, following the synthesis method in~\cite{achouri2014general} and read
\begin{subequations}
\begin{align}
\chi_\text{ee}(x) &= \frac{2}{k_z}\tan{(k_x x)},\\
\chi_\text{mm}(x) &= \frac{2k_z}{k_0^2}\tan{(k_x x)}.
\end{align}
\end{subequations}
It is interesting to note that this specific case of negative refraction leads to purely real susceptibilities and thus passive, lossless and fully efficient refractive metasurfaces. FDFD simulations are used to evaluate the forces acting on these five metasurfaces under Gaussian illumination versus the incidence angle and the results are plotted in Figs.~\ref{Fig:LateralFf}.
We see that the longitudinal forces are zero only when $\theta_\text{i} = \theta_\text{i,spec}$. When the incidence angle deviates from the specified one, then the longitudinal forces are either repulsive or attractive. Similarly, the curves corresponding to the lateral forces in Fig.~\ref{Fig:LateralFfx} cross the dashed black line at the expected values precisely when $\theta_\text{i} = \theta_\text{i,spec}$, except for the metasurface synthesized for $\theta_\text{i,spec} = 75^\circ$. This may be explained by the fact that, for such large angles, undesired scattering occurs in our simulation scheme due to the way the incident wave is numerically generated.

\subsection{In-Plane Rotational Force}

Generating an in-plane rotational force, corresponding to a rotation of the metasurface in the $xy$-plane, may be realized with the transformation of an incident plane wave into a transmitted wave possessing angular momentum. The conservation of angular momentum will result in a rotation of the metasurface in the direction opposite to that of the beam. A common example of such beam, is the Bessel beam with topological charge $m\neq 0$. For simplicity, we consider the transformation of a normally incident plane wave into a normally transmitted TM-polarized ($H_z=0$) Bessel beam of order $m$, as depicted in Fig.~\ref{Fig:F4}. The longitudinal electric field of the Bessel beam is given, in cylindrical coordinates, by~\cite{hernandez2007localized}
\begin{equation}
\label{eq:bessel}
E_z(\rho,\phi) = Ae^{jm\phi}e^{-jk_z z}J_m(k_\rho \rho),
\end{equation}
where $A$ is a complex constant, $m$ is the order of the Bessel beam, $k_z$ and $k_\rho$ are the longitudinal and transverse wavenumbers, respectively. From~\eqref{eq:bessel}, all the field components can be computed and the corresponding rotational force due to the transmitted Bessel beam and which acts on the metasurface is found, by applying~\eqref{eq:TAF}, to be
\begin{equation}
\label{eq:FbeforeInt}
\langle F_\phi \rangle^\text{t} = -\frac{A^2 k_z m \epsilon_0}{2k_\rho^2} \iint_S  \frac{J_m(k_\rho \rho)^2}{\rho}~\rho d\rho d\phi.
\end{equation}
It is possible to obtain a closed form expression of the force by performing the integration in~\eqref{eq:FbeforeInt} over a circular surface. Accordingly, we next assume that the metasurface has a circular shape of radius $r$. In that case, the rotational force due to the transmitted wave is
\begin{equation}
\label{eq:RotForce}
\begin{split}
&\langle F_\phi \rangle^\text{t} = -\frac{A^2k_zm\sqrt{\pi}(k_\rho r)^{2m}\epsilon_0\Gamma\left(m+\frac{1}{2}\right)^2}{\Gamma(m+1)\Gamma(m+\frac{3}{2})\Gamma(2m+1)}\\&
\times_{2}F_3 \left (m+\frac{1}{2}, m+\frac{1}{2}; m+1, m+\frac{3}{2}, 2m + 1;-k_\rho^2 r^2 \right ),
\end{split}
\end{equation}
where $\Gamma(x)$ is the gamma function and $_{2}F_3(a,b;x)$ is a generalized hypergeometric function. Note that because the incident wave is a plane wave, the only contribution to the rotational force is due to the transmitted wave. In~\eqref{eq:RotForce}, the parameter $A$ must be determined so as to satisfy power conservation between the power of the incident plane wave and the power of the transmitted Bessel beam. This is achieved by integrating the $z$-component of the Poynting vector of the Bessel beam over the circular area of the metasurface to find the transmitted power. The parameter $A$ is then found by equalizing the incident power to the transmitted power.

\begin{figure}[h!]
\centering
\includegraphics[width=1\linewidth]{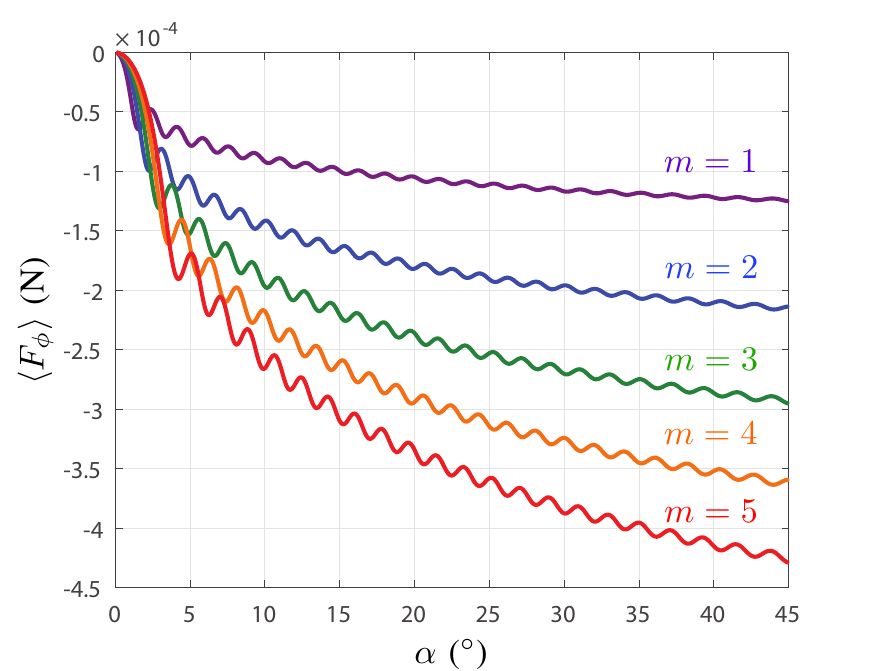}
\caption{Rotational forces versus  cone angle for Bessel beams of different topological charges.}
\label{Fig:BesselF}
\end{figure}
In order to evaluate the rotational force that would be exerted on the metasurface, we have plotted relation~\eqref{eq:RotForce} for $m={1,2,3,4,5}$ versus the Bessel beam cone angle\footnote{This angle is used to define the transverse and longitudinal wavenumbers of the Bessel beam, i.e. $k_\rho = k_0\sin{(\alpha)}$ and $k_z = k_0\cos{(\alpha)}$.}, $\alpha$. The radius of the metasurface is such that the total surface area is the same as that of the rectangular metasurfaces of dimensions $L_x \times L_y$ discussed previously, the radius is thus given by $r=\sqrt{L_xL_y/\pi}$. The corresponding results are plotted in Fig.~\ref{Fig:BesselF}.

As expected, the rotational force is proportional to the topological charge since the latter is directly related to the angular momentum. Due to the complex nature of this transformation, we have not investigated the variation of the rotational force under different incidence angles since it would involve 3D FDFD simulations. More thorough evaluation are thus left for potential future works.

\section{Conclusion}

We have presented different electromagnetic field configuration and metasurface structures that may be used to control radiation pressure and achieve repulsive, attractive, lateral and rotational forces. This work may thus extend the range of motion of solar sail based spacecraft. A potential future direction would be to experimentally verify these relations with acoustic metasurfaces (instead of electromagnetic) so that the amplitude of the forces would be easier to measure.

\section*{Acknowledgment}

This work was accomplished in the framework of the Collaborative Research and Development Project CRDPJ 478303-14 of the Natural Sciences and Engineering Research Council of Canada (NSERC) in partnership with the company Metamaterial Technology Inc.

\bibliographystyle{IEEETran}
\bibliography{NewLib}

\end{document}